\input{aipcheck.tex}

\documentclass[
    ,final            
  ]
  {aipproc}

\layoutstyle{6x9}
\usepackage{amsfonts}
\usepackage{amsmath}
\usepackage{amssymb}

\begin{document}

\title{Towards a New Prescription for the Tidal Capture of Planets, Brown Dwarfs and Stellar Companions}


\author{Niyas Madappatt Alikutty}
{
  address={Department of Physics and Astronomy, Macquarie University, Sydney, NSW 2109, Australia}
}

\author{Orsola De Marco}{
  address={Department of Physics and Astronomy, Macquarie University, Sydney, NSW 2109, Australia}
}

\author{Jason Nordhaus}{
address={Princeton University, Department of Astrophysical Sciences,
         116 Peyton Hall, Princeton, NJ, USA}
}

\author{Mark Wardle}{
  address={Department of Physics and Astronomy, Macquarie University, Sydney, NSW 2109, Australia}
}

\begin{abstract}
We study the evolution of the orbit of a planetary companion as the primary star evolves during the red giant branch phase. Our aim is to determine a prescription for the capture radius and capture timescales for companions in a wide range of masses, from planet to star, including the effect of a non constant $f$ parameter. Our initial results reproduce past findings. According to our study the mass-loss during the red giant branch does not significantly affect the semi major axis of the planet and for a 1.5 $\rm{M}_\odot$ star the capture radius is approximately $300~\rm{R_\odot}$ or 1.4 AU.
\end{abstract}

\maketitle


\section{Introduction}
  The evolution of a star with a planetary companion does not differ a lot from the evolution of a single star until the planetary companion becomes too close to the star due to stellar expansion or orbital evolution. If the companion gets too close to the star it can be tidally captured. The farther out a companion can be captured by an evolving star, the larger the birthrate of stars that derive from binary and planetary interactions (e.g., cataclysmic variable stars). Population synthesis models \citep{politano1996} need to treat tidal capture if they are to make credible predictions of binary population densities and birthrates (e.g., the type \rm{I}a supernovae).
  
Though many studies have been conducted on the subject (e.g., \citealt{Villaver2009}), very few have taken into account the effect of synchronisation on the evolution of the orbital parameters. In the case of planetary companions, synchronisation does not affect the orbital evolution, but when companions get heavier, as in the case of brown dwarfs and low mass main sequence stars, synchronisation can play a major role on the evolution of the orbit. When an object starts to tidally synchronise, it slows the rate of the evolution of the orbital separation. In the absence of mass loss, the orbit will not evolve in a perfectly synchronised system. The long term purpose of this study is to find a general prescription for tides on orbits of planets and  stellar companions that goes beyond what has been done by \cite{soker1996}.
  
In addition, our study will encompass the effect of the dimensionless parameter $f$~\citep{Phinney1992}, which depends on the convective mixing length parameter. For a larger value of $f$ the strength of the tides will be stronger and the evolution of the orbit will be faster. Most studies use $f = 1$ (e.g.,~\citealt{Verbuntphinney1995}, ~\citealt{soker1996}), but ~\cite{Nordhaus2010} has shown that $ f$ need not be constant to match observations of post-main sequence binaries, which means that the functional dependence of $f$ remains unconstrained.  
  
  \section{Evolution of the orbital separation}
  
The evolution of the semi major axis is governed by competing processes. During the red giant branch (RGB) phase of the star, tidal force, gravitational and frictional drag, variation in mass of the star, and changes in the planet mass due to accretion of stellar gas as well as ablation by the star, all play their parts. However, e.g.,  \cite{Villaver2009} established that during the RGB phase, gravitational and frictional drag, along with the mass gained by the planet by accretion do not contribute significantly to the evolution of the semi major axis. Furthermore, in accordance with the study of~\cite{Nordhaus2010}, we assume the mass of the secondary remains constant through the evolution. For this study we consider a $\rm{1~M_{J}}$ planet revolving around a $\rm{1.5~M_{\odot}}$ star,  with  zero eccentricity. 
  
 The orbital change can be expressed as the sum of the change due to mass-loss, and that due to tidal interaction
\begin{equation}
\left(\frac{\dot{a}}{a}\right) ~\approx ~ \frac{\dot{\rm{M}}_{*}}{\rm{M}_{*}}~+~\left( \frac{\dot{a }}{a}\right)_{tide},
\end{equation}
\noindent where $a$ is the semi major axis, M$_{*}$ is the mass of the star and $\dot{a}$ and $\rm{\dot{M}}_{*}$ are their time derivatives (e.g.,~\citealt{Nordhaus2010}).

The tidal force plays a very important role in the evolution of the semi major axis during the RGB phase of the star. For RGB stars, the surface rotation period is much less than the orbital period of the secondary.  In this situation, the sign of the tidal torque is such that the orbit decays. Even though there are various tidal prescriptions discussed (e.g., ~\citealt{hut1981}), the most widely used for convective stars is that formulated by ~\cite{Zahn1966, Zahn1977, Zahn1989}. Tidal decay in this mechanism is governed by convective eddies in the star's envelope:
\begin{equation}
\left(\frac{\dot{a}}{a}\right)_{\rm{tide}}= -\frac{4} {7}~\frac{f}{\tau}~\frac{\rm{M}_{\rm{env}}}{\rm{M}_*} ~q (1+q)\left(\frac{R_*}{a}\right)^8 \left(1- \frac {\Omega}{\omega}\right)
\end{equation}

\noindent where $\omega$ is orbital frequency of the companion, $\Omega$ is spin frequency of the star, $\rm{M}_{env}$ is the mass of the envelope, $\rm{R}_*$ is the radius of the star and $q$ is the mass ratio between the planet and the star. We are also including the effect of synchronisation between the orbital frequency of the companion and the spin frequency of the star, though it plays very little role in the evolution of the orbit for a planetary mass companion. The parameter $f$ is a dimensionless quantity, equal to $1.01 ({\frac{\alpha}{2}})^\frac{4}{3}$, where $\alpha$ is the mixing length parameter~\citep{Verbuntphinney1995}. For the purpose of the current study we have maintained $f$ constant, although it will be varied in the future. The parameter $\tau$ is the convective eddy turnover timescale in the envelope. 

We are using the stellar evolution code, MESA of ~\cite{paxton2010} for the evolution of the star. We evolved a binary system consisting of a $1.5~ \rm{M}_{\odot}$ star and a companion of $1~\rm{M}_{\rm{J}}$ through the entire RGB phase of the star. During each time step equaling 100 years,  we update the value of the stellar parameters and calculate the new value of the $a$ in accordance with ~\cite{Zahn1989}. The parameter $\alpha$ is taken to be 2 in accordance with ~\cite{ZahnBouchet1989}.The initial spin of the star at the begin of the RGB is taken to be 3 km~s$^{-1}$~\citep{deMedeiros1999}. 

\section{Results}

We arbitrarily start our code when the primary has a radius of $5~ \rm{R}_{\odot}$. We run three tests where we vary the initial separation of the $1~\rm{M_J}$ : $275~\rm{R}_\odot, ~320~\rm{R}_\odot,~360~\rm{R}_\odot$. Our preliminary findings are similar to those of ~\cite{Villaver2009} and ~\cite{Nordhaus2010}. We observe that the mass-loss during the RGB does not contribute as much to the evolution of the semi major axis, $a$,  of the binary even though the $1.5~\rm{M}_{\odot}$ star has lost roughly 8 percent of its initial mass. Figure 1 shows the evolution for the semi major axis for entire RGB phase of the star. The increase in $a$ towards the end of the RGB evolution is due to mass-loss. The mass-loss during the AGB will have a much larger effect. In addition, the detailed calculations of the MESA code during the thermally pulsating AGB are likely to result in a different evolution of the orbital separation compared to tidal codes that use off the shelf stellar evolutionary tracks (e.g.,~\citealt{Nordhaus2010}). Figure 2 just focuses on the last 200,000 years of RGB evolution. The two plots show a range of initial separations for which the companion may get captured or evade capture. The thick continuous line shows the evolution of the primary's radius, the three discontinuous lines describe the evolution of the semi major axis $a$. As the system evolves we find that the distance beyond which the companion does not get captured is approximately 320 $\rm{R}_{\odot}$ or 1.5 AU. This is approximately 2.5 times the maximum RGB radius of this star. Further study needs to be done in the cases where the companion is in the range 300 - 320 $\rm{R}_{\odot}$, here the companions may get captured during the beginning of the horizontal branch phase of the star. 

\begin{figure}
  \includegraphics[height=82mm]  {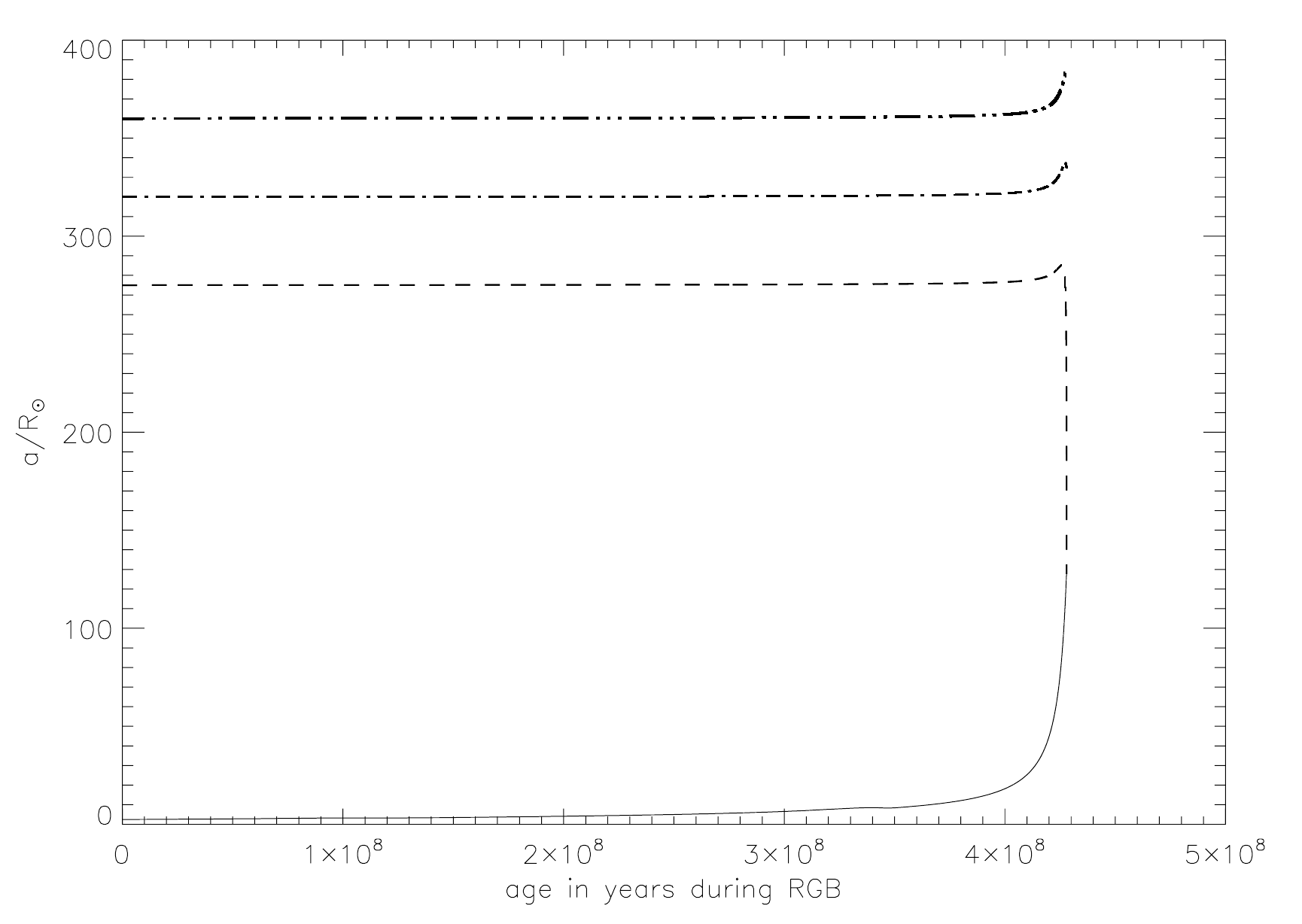}
  \caption{Evolution of the radius of the star during the entire RGB phase (continuous line) and the orbital evolution of the semi major axis with three initial separations (275, 320, 360 $\rm{R_\odot}$, discontinuous lines).}
\end{figure}
\begin{figure}
  \includegraphics[height=82mm]  {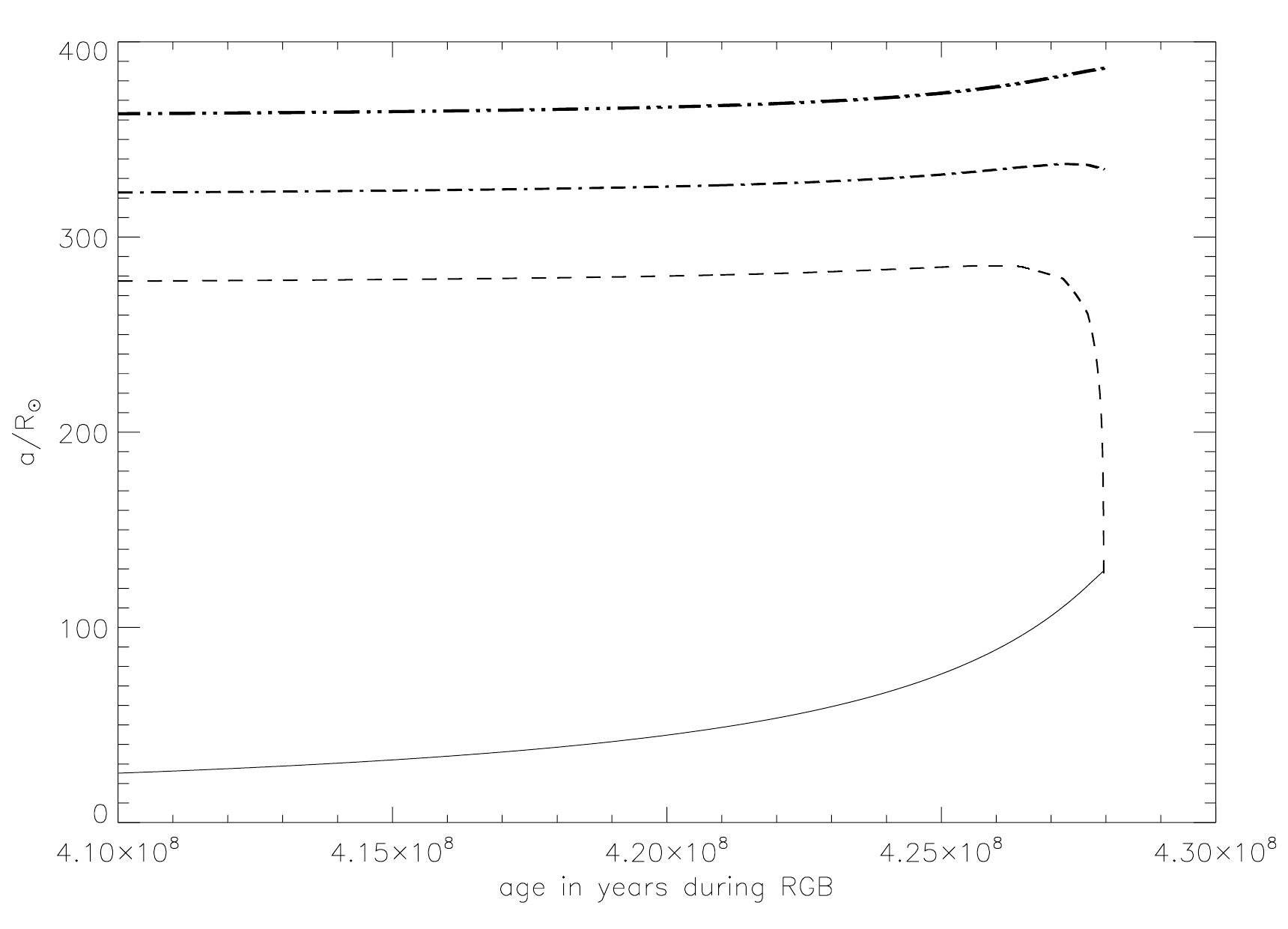}
  \caption{Evolution of the radius of the star during the last  200,000 years of the RGB (continuous line) and of the orbital evolution of the semi major axis for 3 distinct initial separations during the same period (275, 320, 360 $\rm{R_\odot}$, discontinuous lines).}
\end{figure}
\subsection{Future Work}
We expect to find that since mass-loss dominates during the AGB phase of the star, it is unlikely that the companion will get engulfed if it is much farther than a couple of stellar radii. This was also the result of \cite{Nordhaus2010}. We will extend the work of \cite{Nordhaus2010} to more massive companions, including the effect of synchronisation. By including a variable $f$ value we will extend the results of \cite{Villaver2009}, and by including a proper treatment of the thermally pulsating AGB we will have a better representation of the AGB star radial evolution, determining when and how companion get engulfed during this phase. 


\begin{theacknowledgments}
We thank Bill Paxton for the calculation of the stellar evolutionary track. NM also acknowledges funding from Macquarie University Research Excellence PhD scholarship and thanks the meeting's organisers for financial support.

\end{theacknowledgments}



\bibliographystyle{aipproc}   


\hyphenation{Post-Script Sprin-ger}

\end{document}